\documentclass{ws-ijmpa}

\newcommand{\jp}{J/\psi}
\newcommand{\TeV}{\ensuremath{\mathrm{Te\kern -0.1em V}}}
\newcommand{\MeV}{\ensuremath{\mathrm{Me\kern -0.1em V}}}
\newcommand{\MeVCSq}{\ensuremath{\MeV\!/c^2}}
\newcommand{\nbinv} {{\mathrm{nb}^{-1}}}
\newcommand{\pbinv} {{\mathrm{pb}^{-1}}}
\newcommand{\GeV}{\ensuremath{\mathrm{Ge\kern -0.1em V}}}
\newcommand{\GeVC}{\ensuremath{\GeV\!/c}}
\newcommand{\pt}{\ensuremath{p_{\rm T}}}
\newcommand{\rg}{\rightarrow}

\begin{document}

\markboth{Giulia Manca, LHCb Collaboration}
{Studies of Open Charm and Charmonium Results from LHCb}

%
\catchline{}{}{}{}{}
%

\title{Studies of Open Charm and Charmonium Production at LHCb}

\author{Giulia Manca\footnote{Giulia.Manca@cern.ch} (on behalf of the LHCb Collaboration)}

\address{Department of Physics, University of Cagliari, 
Cittadella Universitaria di Monserrato, 09042 Monserrato, Cagliari, Italy
}

\maketitle


\begin{abstract}
We present recent results from charmonium
and open charm production at the LHCb experiment at CERN, Geneva.
We concentrate on studies for the measurement of the 
cross section $pp \rightarrow \jp X$ in its decay channel with
two muons, showing the agreement of the simulation with the
data in few key distributions.
We also show the reconstructed 
modes of 
$D^0 \rightarrow K^-\pi^+$,
$D^0 \rightarrow K^-\pi^+\pi^0$, 
$\Lambda_c^+ \rightarrow p K^- \pi^+$ and 
$D^+ \rg K_s^0 \pi^+$ and their~prospects.

\end{abstract}

\ccode{PACS numbers 11.25.Hf, 123.1K}

\section{LHC and LHCb}

After a troubled start-up, the Large Hadron Collider 
(LHC) performed the first proton-proton collisions at a 
center of mass energy of 7~$\TeV$ in March 2010.
Since then the accelerator has been delivering
data with increasing stability and intensity, 
running at a steady value of instantaneous 
luminosity L of 1-2$\cdot 10^{29} \mathrm{cm^{-2}s^{-1}}$
at the time of writing.
So far the LHC has delivered an integrated luminosity $\cal L$ of 
16~$\nbinv$ of data on tape for each experiment.
The LHCb experiment\cite{lhcb} is a forward spectrometer 
optimised to study the physics of $B$ mesons. 
The geometry of the detector is such to 
maximise the acceptance for $B$'s, which 
are mainly produced in the forward region.
The detector consists of a high precision vertex 
detector (VELO), a first Ring Imaging Cerenkov detector (RICH), RICH-1,
two tracking systems located 
before (TT) and after (IT,OT) the dipole magnet (B=1.4 T),
a RICH-2, the electromagnetic and hadronic calorimeters, 
and the five muon stations (M1-M5).
The silicon planes of the VELO are retracted 
during injection, but approach to within 8 mm 
of the beam during stable operation.
The trigger consists of three levels: 
L0, which is purely hardware, HLT1 and HLT2, which are purely software based,
with the purpose of reducing the rate from 40 MHz to 2 kHz.
At the time of writing, the interaction rate is sufficiently low that only L0 and HLT1 
are required.

\section{$J/\psi$ Results}

\begin{figure}[pb]
 \centering
 \includegraphics[width=5.5cm]{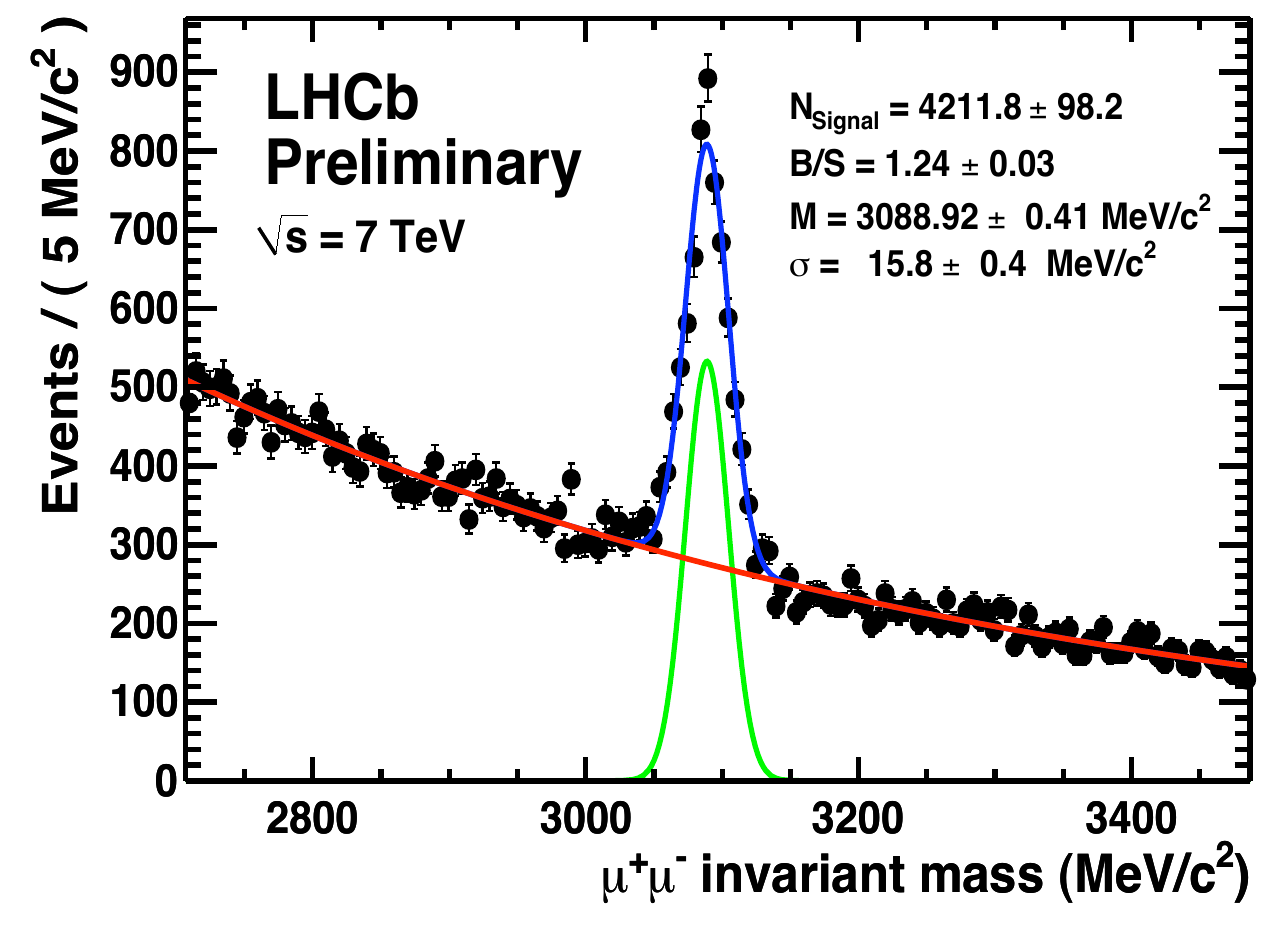}
 \includegraphics[width=5.5cm]{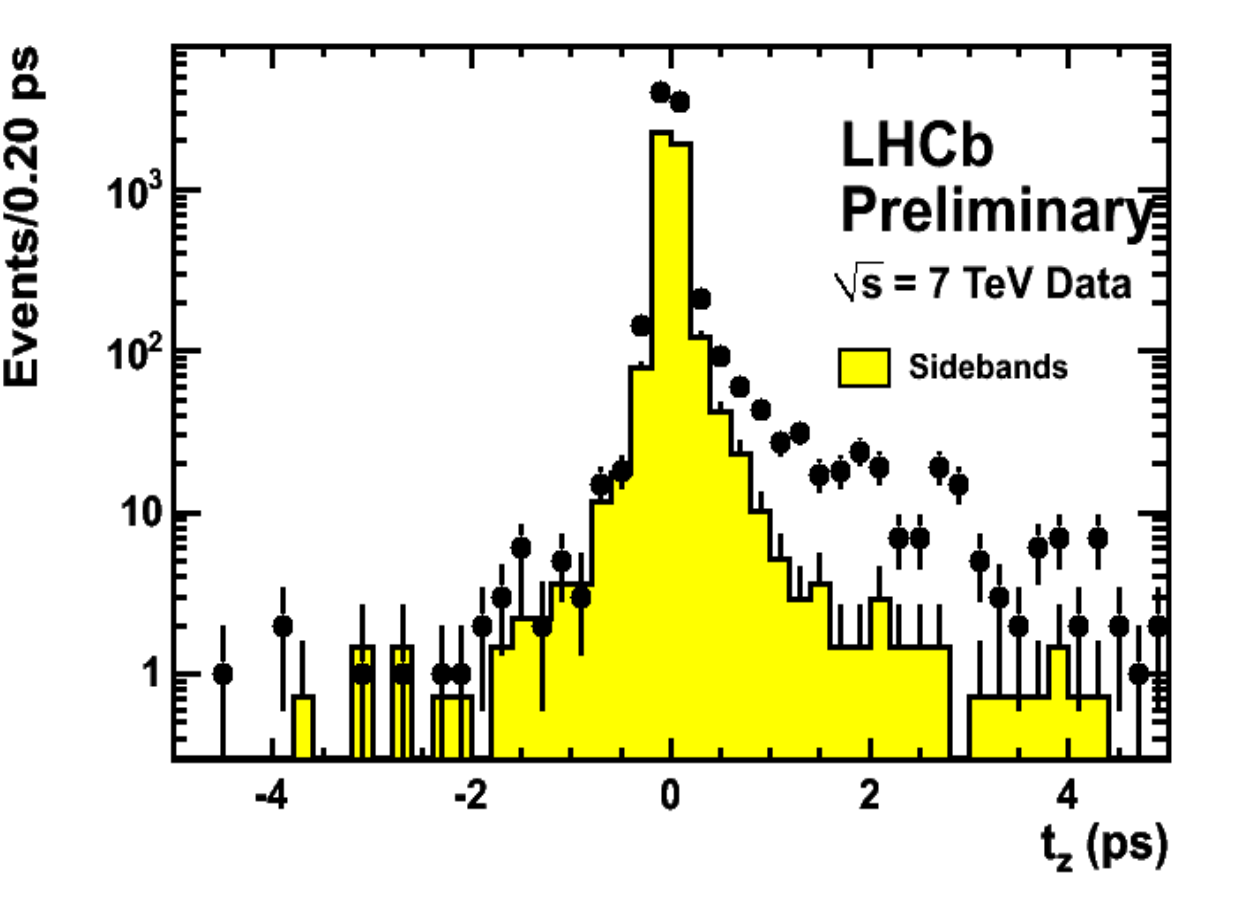}
 \caption{{\it Left}: Invariant mass distribution of $\mu^+\mu^-$. {\it Right:}
Pseudo-proper time distribution for $\jp$ candidates. The enhancement at $t_z\gtrsim$~3~ps
is due to the presence of $\jp$'s from $B$ decays.
 \label{fig:jpsimass}}
\end{figure}

Despite its first observation in 1974,
the production of  $J/\psi$ particles at hadron colliders 
is still unclear.
Several models have tried to reproduce the behaviour
of the data collected e.g. by CDF-II at Fermilab\cite{cdf}; among these,
NRQCD\cite{nrqcd} is one of the most successful ones in reproducing
the $p\bar{p}\rightarrow J/\psi$ cross section behaviour, but it 
fails to predict e.g. the polarisation.
The data from the LHC experiments will be crucial
in solving this puzzle. For LHCb, the $J/\psi$ cross section 
measurement is a fundamental milestone towards the measurement
of the $B$ cross section and CP-violation physics,  
as well as an important test of the detector performance.\\
$\jp$'s are produced at the LHC\cite{jpsi} in three main ways, 
directly from $pp$ interaction ({\it ``prompt''}), or indirectly through 
the decay of an intermediate state such as $\chi_c$ ({\it ``indirect''}) or 
from a $B$ hadron ({\it ``delayed''}).
The cross section for $\jp$ prompt production is 
of the order of 10~$\mu$b and for the delayed $\jp$ about one order
of magnitude smaller. The branching ratio of the $\jp\rightarrow\mu\mu$
is $\sim 6\%$.
The muon trigger efficiency at LHCb is well above 90\%
.
The muon reconstruction efficiency has been estimated 
using $\jp$ events themselves  as a function of the $\mu$ $\pt$
and it is overall (97.3$\pm$1.2)\% for $\pt>6~\GeVC$.
The invariant mass distribution of $\mu^+\mu^-$ is 
shown in Fig.~\ref{fig:jpsimass}({\it left}) for $\cal L$=14~$\nbinv$.
Around 4200 signal events are observed.
The mass resolution is about 16~$\MeVCSq$, about 20\% higher than the Monte Carlo (MC)
predictions. This is expected to improve as the alignment of the detector is better
understood.
A ``pseudo-proper time'', $t_z$, of the $\jp$ can be reconstructed, defined 
as $t_z=(\Delta z/p_z)\cdot M_{\jp}$,
where $\Delta z$ is the distance between the primary and secondary vertex, 
$p_z$ is the component of the $\jp$ momentum along the $z$ axis,
 $M_{\jp}$ is the $\jp$ mass. $t_z$ is useful in distinguishing
between delayed and prompt plus indirect $\jp$ production.
The $t_z$ distribution in the data
is shown in Fig.~\ref{fig:jpsimass}({\it right}). 
The enhancement at positive lifetimes is indeed an indication of 
the presence of $\jp$ from $B$ in the sample. 
The acceptance for $\jp$ particles has been studied with unpolarised Monte 
Carlo 
samples produced with Pythia 6.4 and found to be about 13\%.
We plan to measure the $\jp$ production cross section in five bins 
of rapidity $y$ and seven bins of 
 $\pt$ with $\cal{L}\sim$10~$\pbinv$. The $\pt$ and $y$ distributions 
in the data agree reasonably with the MC ones.
The effect of the polarisation on the 
measurement has been studied in the MC  and 
ranges between 5 and 25\% depending on $y$ and $\pt$.

\section{Open Charm Results}

Reconstruction of decays of $D$ mesons (``open charm'') is pivotal
to crucial measurements such as $D^0$-$\bar{D^0}$ mixing or violation of
Charge-Parity(CP) conservation.
The $c\bar{c}$ production cross section for $pp$ collisions at 
$\sqrt{s}=7~\TeV$ is of the order
of 5~mb, roughly ten times larger than the $b\bar{b}$ cross section at the same 
energy. Charm particles are produced both prompt and from $B$ decays.
At LHCb 
the trigger requires, in nominal conditions,  energy thresholds
typical of $B$ events which results in low efficiency for triggering on 
charm particles.
However, in this initial phase of data taking with collision 
rates below 25~kHz, we have been able to lower the trigger 
requirements to reach efficiencies $\simeq 40$-$50\%$ for promptly produced charm particles.\\
In 14~$\nbinv$ of data we observe 6340 (1101) $D^0$'s reconstructed through 
their decays into $K^-\pi^+$ in untagged (tagged from the $D^*$) mode.
The high statistics of these decays make them the ideal candidates for
measurements like cross sections and  $D^0$-$\bar{D^0}$ mixing.
In addition, LHCb has already managed to reconstruct decays like 
$D^0 \rightarrow K^-\pi^+\pi^0$, 
$\Lambda_c^+ \rightarrow p K^- \pi^+$ and 
$D^+ \rg K_s^0 \pi^+$.
The difficulty of these channels  is mainly in the experimental 
techniques involved in the identification and reconstruction of the
 decay products,
particularly challenging in a busy environment
like a hadron collider.
The $D^0\rightarrow K^-\pi^+\pi^0$ demonstrates high reconstruction efficiency 
for neutral particles, and the good functioning of the RICH used for kaon identification; 
the observation of $\Lambda_c^+\rightarrow p K^- \pi^+$
is a proof of the good vertex resolution of the VELO and $D^+\rightarrow K_s^0 \pi^+$ 
shows the ability of reconstruct particles also with just
one track close to the vertex.

\section{Conclusions and Outlook}
The charm program at LHCb has just started and already looks extremely 
promising.
With all the tools into place we have set the basis for the 
$\jp$ cross section measurement, 
which is a crucial step in the understanding 
of the detector 
as well as an important physics result which could solve existing 
open issues in QCD.
We have observed several open charm decay modes 
such as $D^0\rightarrow K^-\pi^+$, important
for the $D^{0}$-$\bar{D^{0}}$ mixing measurement, which we
expect to produce the world-best measurement with 
100~$\pbinv$ of data.
With the same luminosity we also expect to set competitive limits
on rare charm decays like $D^{0}\rightarrow \mu^+\mu^-$.

%


\end{document}